\documentclass[12pt, twoside]{article}
\usepackage{amsmath,amsthm,amssymb}
\usepackage{rotating}
\usepackage{times}
\usepackage{enumerate}
\usepackage{subfigure}
\usepackage{color}
\usepackage[normalem]{ulem}

\usepackage{graphicx}
\graphicspath{{../pdf/}{D:\ImagesforProjectLatex}}

\theoremstyle{definition}

\numberwithin{equation}{section}

\newcommand{\ur}{U_\rho}
\newcommand{\wwr}{W_\rho}
\newcommand{\net}{{Notbohm \emph{et al.}  \cite{notb} }}
\newcommand{\zr}{Z_\rho}

\newcommand{\zrsq}{Z_{\sqrt{\rho}}}
\newcommand{\zrm}{Z_{\rho^{-1}}}

\newcommand{\be}{\begin{equation}}
\newcommand{\ee}{\end{equation}}

\newcommand{\e}{\varepsilon}
\newcommand{\er}{\varepsilon_r}
\newcommand{\et}{\varepsilon_\theta}
\newcommand{\ef}{\varepsilon_\varphi}
\newcommand{\sfi}{\sigma_\varphi}\newcommand{\ga}{\alpha}\newcommand{\gb}{\beta}
\newcommand{\xm}{{\xi_-}}
\newcommand{\xp}{{\xi_+}}
\newcommand{\s}{\sigma}
\newcommand{\sr}{\sigma_r}
\newcommand{\st}{\sigma_\theta}

\newcommand{\RR}{\mathbb{R}}

\newcommand{\bu}{{\boldsymbol{u}}}

\newcommand{\bm}{{\boldsymbol{m}}}

\newcommand{\bE}{{\boldsymbol{E}}}
\newcommand{\bS}{{\boldsymbol{S}}}

\newcommand{\bv}{{\boldsymbol{v}}}

\newcommand{\bI}{{\boldsymbol{1}}}

\newcommand{\bx}{{\boldsymbol{x}}}
\newcommand{\p} {\partial}

\newcommand{\tr}{{\rm tr }}

\def\XXint#1#2#3{{\setbox0=\hbox{$#1{#2#3}{\int}$}
     \vcenter{\hbox{$#2#3$}}\kern-.5\wd0}}

\begin{document}


\baselineskip=19pt

\title{A Model for Compression-Weakening Materials and the Elastic Fields due to Contractile Cells}

\author{
Phoebus Rosakis\\
Department of  Theoretical and Applied Mathematics\\ University of Crete, Heraklion 70013, Greece\\
rosakis@uoc.gr\\   \\ Jacob Notbohm\\Department of Engineering Physics, University of Wisconsin\\ Madison, WI 53706, USA\\ {jknotbohm@wisc.edu}
\\  \\Guruswami Ravichandran\\
Division of Engineering and Applied Science\\ California Institute of Technology, Pasadena CA 91125, USA\\ravi@caltech.edu
}

\date{}

\maketitle

\begin{abstract}\noindent We construct a homogeneous, nonlinear elastic constitutive law, that models aspects of the mechanical behavior of inhomogeneous fibrin networks. Fibers in such networks buckle when in compression. We model this as a loss of stiffness in compression in the stress-strain relations of the homogeneous constitutive model. Problems that model a contracting biological cell in a finite matrix are solved. It is found that matrix displacements and stresses induced by cell contraction decay slower (with distance from the cell) in a compression weakening material, than linear elasticity would predict. This points toward a mechanism for long-range cell mechanosensing. In contrast, an expanding cell would induce displacements that decay faster than in a linear elastic matrix. 


\end{abstract}

\section{Introduction}\label{intro}

Biological cells can sense the mechanical state of the surrounding extracellular matrix, such as stiffness \cite{2discher}, deformations, forces, or stress {}{\cite{5lo,6reinhart,9winer,shi2014,he}}. This is known as mechanosensing \cite{1vogel}. At the same time, cells actively contract, thereby applying tractions on the extracellular matrix and  deforming it.  The resulting displacement or stress fields  can serve as signals to other cells {\cite{6reinhart,9winer,shi2014}}, thus enabling neighboring cells to detect each  other  \cite{notbth,notb}.
 
Experiments using digital volume correlation with confocal microscopy \cite{26franck} measured displacements in a 3D fibrin matrix caused by contractile fibroblasts seeded in it \cite{notbth,notb}. The fibrin matrix is not a homogeneous material, but rather a random network of slender fibers. The matrix displacements induced by cell contraction were found to decay much slower with distance from the cell than  linear elasticity would predict. Thus, contractile cells  embedded within a fibrin matrix can detect mechanical fields induced by each other at larger distances, compared to cells in a homogeneous gel matrix that behaves like a linear elastic material,  {}{or  cells on a linear elastic substrate where displacements decay even faster (exponentially \cite{he})}. This observation of long-range cell--cell mechanical communication in 3D agrees with previous experiments that showed a similar effect for cells on a 2D {}{fibrin} substrate \cite{9winer,25rudnicki}. The mechanism for the long-range mechanosensing stems from the mechanical behavior of the fibrous network. It was shown \cite{notbth,notb} that  the displacements due to a contracting inclusion in a fiber network decay slower than in a homogeneous linear elastic material, because  fibers lose stiffness in compression. The stiffness loss is due to \emph{microbuckling}, namely buckling of  individual fibers in the network that are in compression. See \cite{Lakes,Kim,kim2} for various aspects of microbuckling.  
\begin{figure}[]\label{fig1}
\centering
\subfigure[]{
  \includegraphics[width=0.45\textwidth]{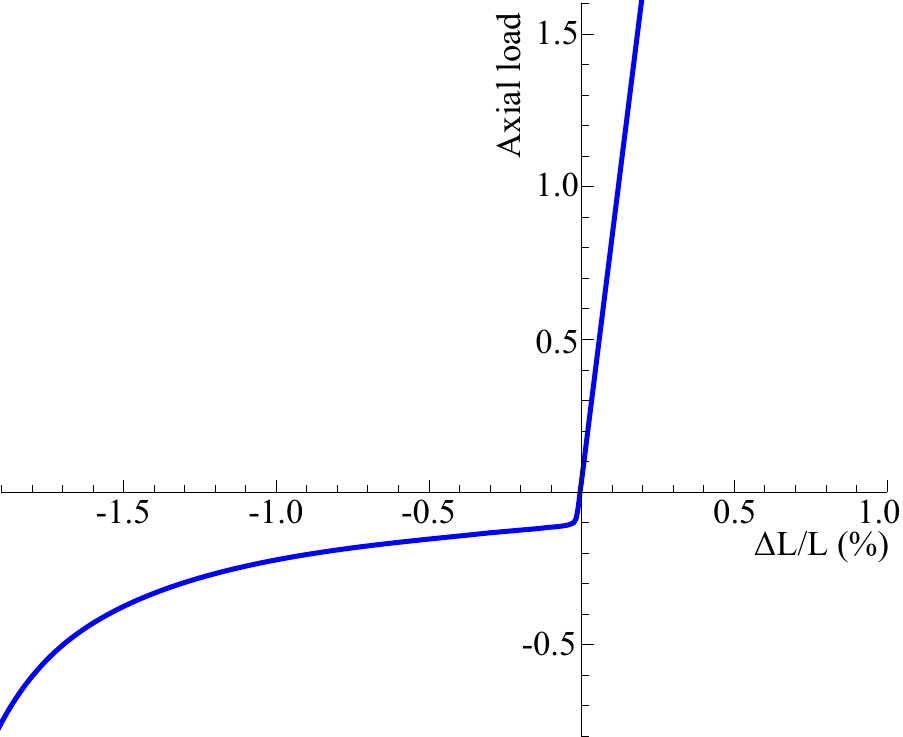}
  \label{subfig1a}}
\quad
\subfigure[]{%
  \includegraphics[width=0.45\textwidth]{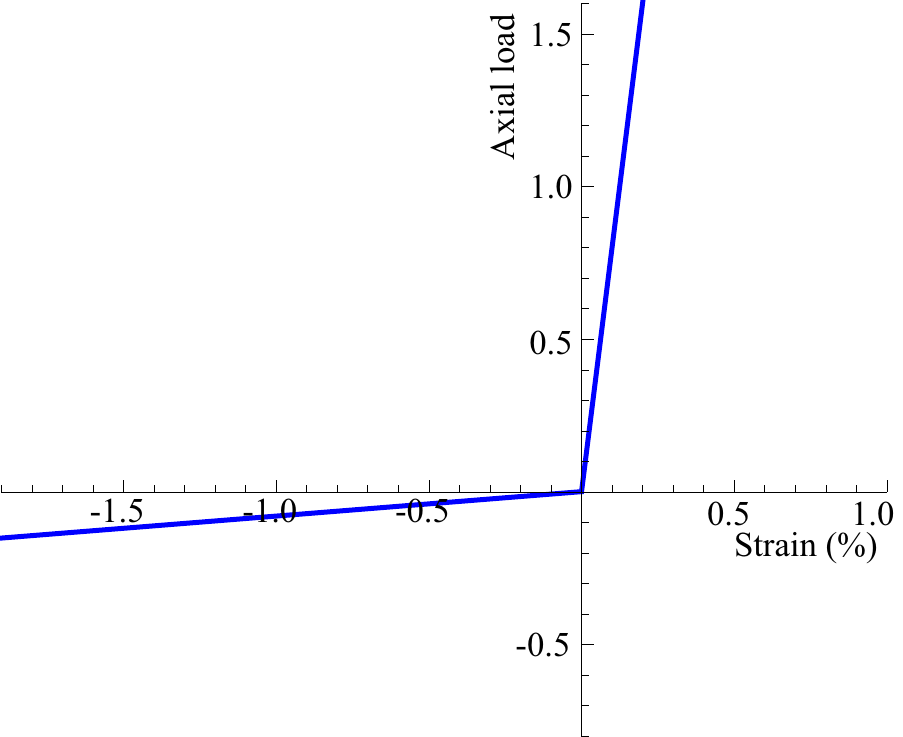}
  \label{subfig1b}}
\caption{
 {\bf(a)} Typical relation between axial load (vertical axis, arbitrary units) and {fractional change in the distance between endpoints} (horizontal axis, percent) of an elastic beam that can buckle. { \bf(b)} One-dimensional piecewise-linear stress-strain curve for a material that weakens in compression. Here $\rho=0.1$.  Horizontal axis: strain $\e$ in percent. Vertical  axis: normalized stress $\s/\ga$, where $\ga$ is a one-dimensional elastic constant.  }
\label{fig:figure}
\end{figure}

We previously developed a finite element  fiber network model \cite{notbth,notb} that treated individual fibers as elements whose force-extension curve has smaller slope in compression than in tension, as in  Fig.~\ref{subfig1b}.  This is an idealization of the  typical relation between axial load  and fractional change in the distance between endpoints  of an elastic beam that can buckle, shown in Fig.~\ref{subfig1a}.  One notes the abrupt change of stiffness that occurs at a negative value of the load (the buckling load) in Fig.~\ref{subfig1a}.  The magnitude of the buckling load depends on the bending stiffness of fibers.  For fibrin,  the bending stiffness has been found to be nearly two orders of magnitude less than the value predicted by the pure bending model of linear elasticity \cite{piech}. Accordingly, the buckling load is essentially taken to vanish in Fig.~\ref{subfig1b}. Simulations of our  finite element  model  in 2D \cite{notbth,notb} with fiber elements obeying the compression weakening stress-strain law of Fig.~\ref{subfig1b}  show that  matrix displacements induced by a contracting spherical inclusion (representing the cell) decay according to a power law $u\sim r^{-n}$ with distance $r$ from the inclusion center.  Values of $n$ depend on the connectivity of the network, but are always in the range $0.2-0.5$, far below the value $n=1$ that 2D linear elasticity would predict.  In 3D, values of $n$ from simulations were in the range $0.6-0.9$, and experiments yielded $n=0.52$, again much  less than the linear elastic value $n=2$.
Significantly, when the microbuckling elements  were replaced by linear elastic ones that do not buckle (same stiffness in compression as in tension)  the network simulations yielded values of $n$ close to the linear elastic predictions \cite{notb}. 

These finite element simulations of  discrete network models  provided strong evidence for the hypothesis formulated  by Notbohm \cite{notbth} and coworkers \cite{notb}: \emph{microbuckling of fibrin enables long-range cell-induced displacements that facilitate mechanosensing.}
The purpose of this paper is to provide theoretical support for this  conclusion. So far,  the evidence comes from  experiments and numerical simulations of a discrete network  \cite{notbth,notb}.  Here we conjecture that  a homogeneous  solid  with  \emph{lower stiffness in compression than in tension} will also exhibit  slower decay of displacements/stresses due to a contracting inclusion, than a linear elastic solid. In other words, the dominant factor responsible for the slow displacement decay is loss of stiffness in compression, rather than  the discrete character of the fiber network. We  show this by first constructing a constitutive model with the requisite properties, then solving some relevant boundary value problems analytically. 

Constitutive modeling is described in detail in Section \ref{sec2}.  The analysis of relevant boundary value problems is spelled out in Sections \ref{sec3},  \ref{sec4}.   Sections \ref{sec2},  \ref{sec3} and \ref{sec4} contain considerable technical details and may be omitted at  first reading.  We  briefly present our main results in Section \ref{sec1} (Results) and  discuss  their significance in Section \ref{disc} (Discussion).  
\section{Constitutive Model}\label{sec2}

\subsection{2D Constitutive Law}\label{C2}

We now describe a special elastic constitutive law, which loses stiffness in compression in a sense to be made precise. Such a constitutive law cannot be linear, even  in the context of small deformations, if it is going to exhibit behavior analogous to that of Fig.~\ref{subfig1b} in more than one dimension.  For simplicity  we consider small deformations (linearized kinematics). The constitutive law itself is nonlinear; it is isotropic, while the principal stresses are piecewise linear functions of the principal strains (eigenvalues of the infinitesimal strain tensor). This last property allows us to solve some problems of interest analytically.

To begin with, suppose the matrix plus the cell together occupy the whole 2D space $\RR^2$ and is composed of linear elastic homogeneous isotropic material (undergoing small deformations, so that the linearized theory of elasticity is used). We thus have a displacement field ${\bu}\colon\RR^2\to\RR^2$. The components of the  (infinitesimal) strain tensor are
$$
\bE=\frac{1}{2}(\nabla\bu+\nabla\bu^T),\quad 
E_{ij}=\frac{1}{2}\left(\frac{\p u_i}{\p x_j}+\frac{\p u_j}{\p x_i}\right).$$
In the matrix except the cell, the  stress tensor is related to the   strain tensor by
\be\label{s1}\bS  =\lambda(\tr\bE)\bI+2\mu\bE.\ee
Here $\lambda$ and $\mu$ are the Lam\'e constants and $\bI$ the identity tensor.  In components, the above reads\footnote{The Einstein summation convention is used: summation over repeated indices is implied, unless indicated otherwise} 
 $$S_{ij}=\lambda E_{kk}\delta_{ij}+2\mu E_{ij}.$$
 The principal stresses $\s_i$ (the eigenvalues of the stress tensor) are related  to the principal strains $\e_i$ (the eigenvalues of the strain tensor) through
\[\s_i=C_{ij}\e_j,\]
where 
\[    C=       \begin{pmatrix} 
      \ga & \gb \\
     \gb & \ga\\
   \end{pmatrix}, \quad \ga=\lambda+2\mu,\quad \gb=\lambda,\]
and $\lambda$, $\mu$ are the Lam\'e  Moduli (elastic constants; $\mu$ is the shear modulus).   In other words, the linear elastic principal stress-strain relations are
\be\label{linel}\s_1=\ga\e_1+\gb\e_2,  \quad \s_2=\gb\e_1+\ga\e_2.
\ee
Positive definiteness of $C$ is equivalent to
 \be\label{pd} \ga>|\gb|.\ee

Our first attempt toward constructing a constitutive law that weakens in compression is to consider piecewise-linear stress-strain relations.  Consider the function 
 \be\label{zr}\zr(x)  =\begin{cases} 
x,&  x\ge 0,
\\  
\rho x,&x<0,\end{cases}\ee
where $0\le\rho\le1$ is the constant \emph{compression stiffness ratio.}  The graph of $\zr$ is the curve in Fig.~\ref{subfig1b}. Note that the function $\zr$ is not linear, but it is  \emph{positive-homogeneous of degree one}, i.e., $\zr(\ga x)=\ga\zr(x)$ for any $\ga>0$ (and any real $x$).

In 1D, one might replace the linear stress-strain relation $\s=\ga\e$, where $\ga$ is a modulus, by the  piecewise linear stress-strain relation $\s=\zr(\ga\e)$; see Fig.~\ref{subfig1b}.  By analogy, in our first attempt, we replace \eqref{linel} with
\begin{align}\label{one}\s_1=&\zr(\ga\e_1+\gb\e_2),  \\
\label{two}\
\s_2=&\zr(\gb\e_1+\ga\e_2).
\end{align}
In effect  this  multiplies stiffness by $\rho$ whenever the corresponding principal stress is negative (compressive).  However, this turns out to be problematic.  Suppose the argument of $\zr$ in \eqref{one} is positive and the argument of $\zr$ in \eqref{two} is negative. Then we have $\p \s_1/\p \e_2=\gb$  while $\p \s_2/\p \e_1=\rho\gb$, which means that if $\rho\not=1$, there is no strain energy density function $W(\e_1,\e_2)$ such that $\s_i=\p W/\p\e_i$. Thus the stress-strain relations \eqref{one}, \eqref{two} are not hyperelastic, except in the trivial case of no weakening ($\rho=1$) which coincides with linear elasticity. Therefore  this model is not satisfactory.

\begin{figure}[]\label{fig2}
\centering
\subfigure[]{
  \includegraphics[width=0.45\textwidth]{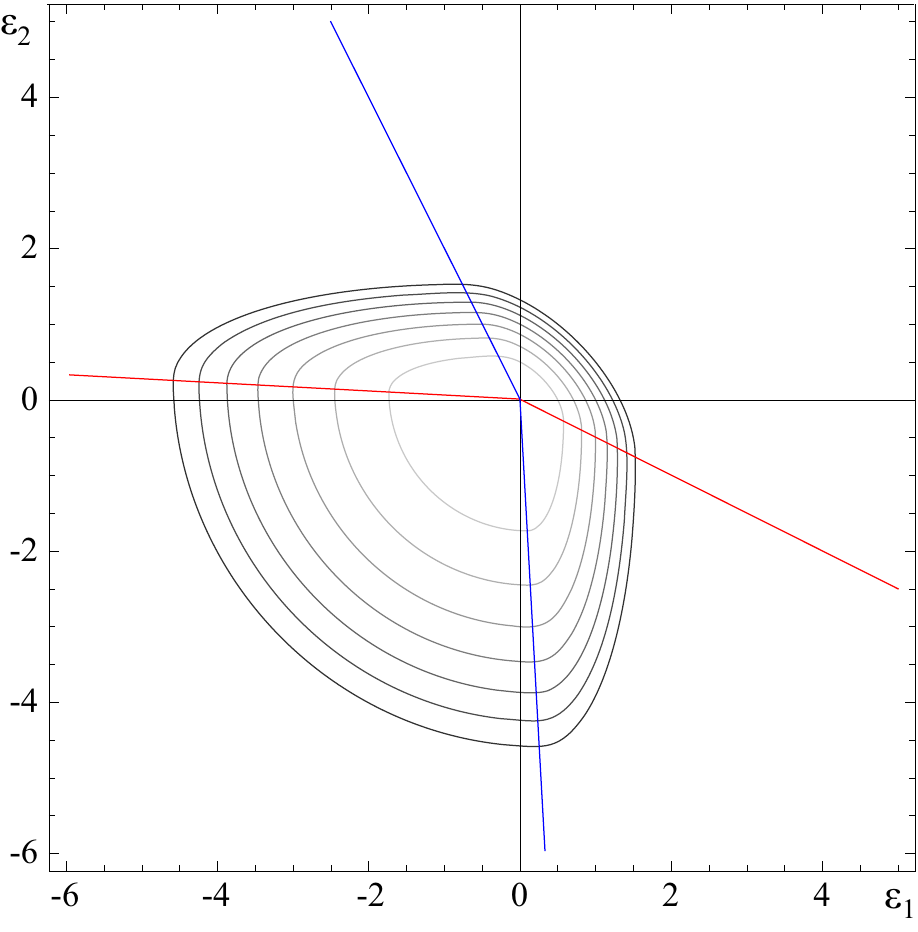}
  \label{fig:subfigure1}}
\quad
\subfigure[]{%
  \includegraphics[width=0.45\textwidth]{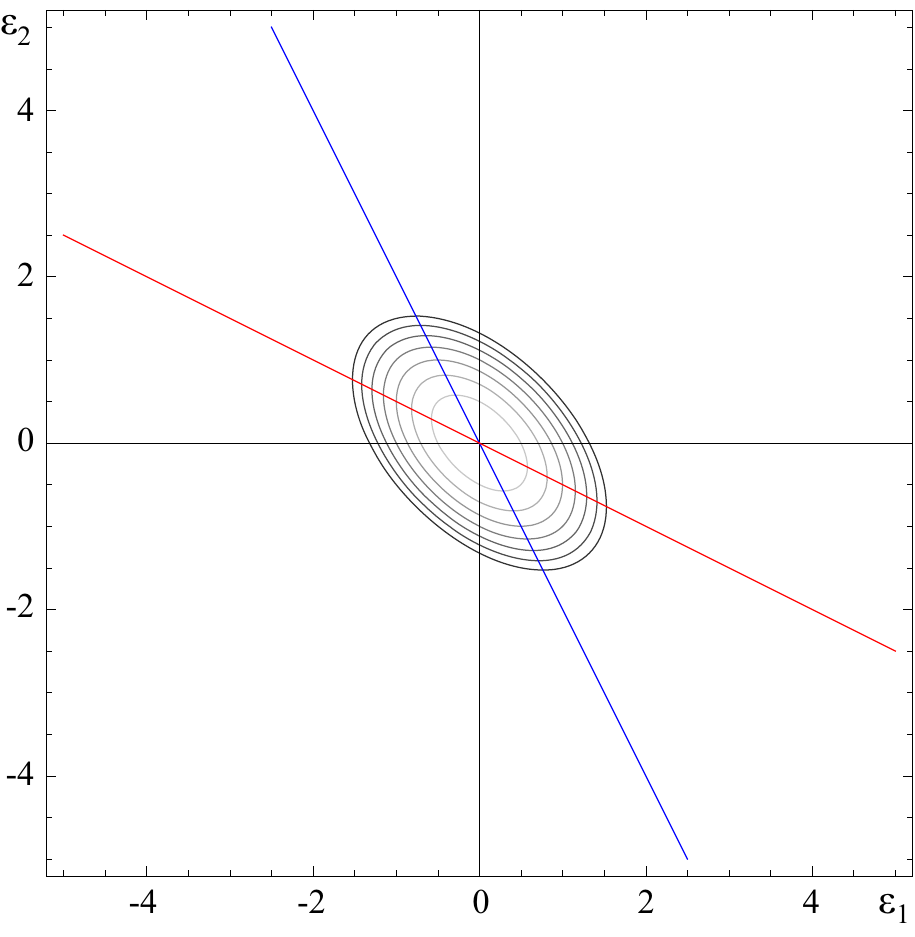}
  \label{fig:subfigure3}}
  
  \subfigure[]{%
  \includegraphics[width=0.45\textwidth]{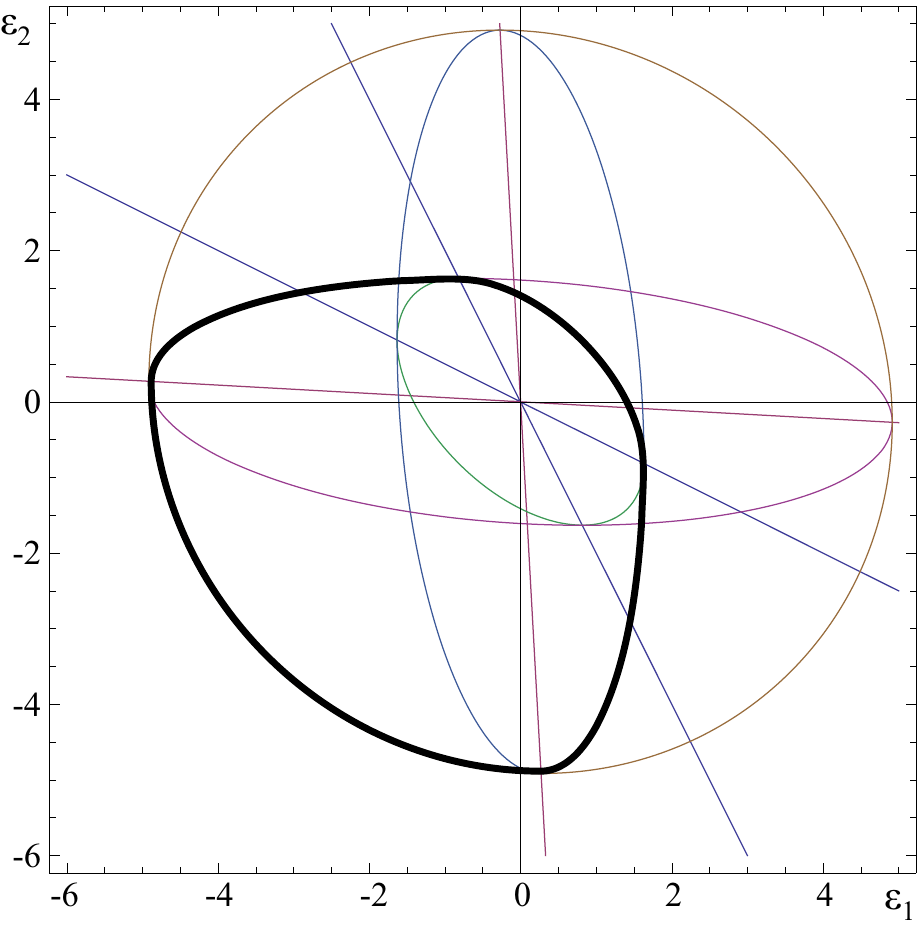}
  \label{fig:subfigure2}}
\caption {{\bf(a)} Level curves of the strain energy density $\wwr$ in the $(\e_1,\e_2)$ plane. {}{Here $\lambda=2\mu$ and $\rho=1/9$; values of $\wwr$ on the level sets are $10^{-4}\times k\lambda$ for $k=1,2,\ldots,7$. Values of strain are in percent}. The blue  and red  straight  lines separate the plane into four sectors. The strain energy density is equal to a different quadratic function in each sector. The blue line corresponds to uniaxial stress, $\s_1=0$. The red line  corresponds to uniaxial stress, $\s_2=0$.  { \bf(b)} For comparison,  level curves of the corresponding linear elastic energy ($\rho=1$).  Blue and red lines correspond to uniaxial stress. { \bf(c)}  A single level curve of the strain energy density (thick curve). The four ellipses (blue, green, purple and brown) are the level curves of the  four quadratic functions that equal the strain energy in different sectors. Switching between these four branches occurs at points on the uniaxial stress lines; these straight lines are where the principal stresses change signs.}

\label{fig:figure}
\end{figure}

 In order to overcome the lack of hyperelasticity just encountered, one might attempt to construct the strain energy function directly.  {}{This is somewhat difficult, however,} since the change in stiffness is supposed to occur when stresses, not strains, change sign. It is more natural to construct the \emph{complementary energy density}  {}{(a function of  stress)}  $U(\s_1,\s_2)$, with the property that 
 \be\label{ues} \frac{\p U(\s_1,\s_2) }{\p\s_i}=\e_i.\ee
Assuming that $W(\e_1,\e_2)$ is strictly convex and continuously differentiable, the stress-strain relations are invertible to the from $\e_i=\hat\e_i (\s_1,\s_2)$  and one has
\[ U(\s_1,\s_2)=\s_i\e_i-W(\e_1,\e_2),\quad \e_i=\hat\e_i (\s_1,\s_2).\]
In 1D, one  might adopt the stress-strain relation $\s=\ga\zr(\e)$, where $\ga$ is a modulus.  Then since $\s=W'(\e)$, we have
\[W(\e)= \frac{\ga}{2}\zrsq^2(\e).\]
Also since the strain-stress relation is $\e=\ga^{-1}\zrm(\s)$, the complementary energy is
\[U(\s)= \frac{\kappa}{2}Z_d^2(\s),   \quad d=1/\sqrt{\rho},\quad \kappa=1/\ga.\]
The complementary energy is quadratic in $Z_d(\s)$, {}{hence piecewise quadratic in $\s$}, while for linear elasticity ($d=1$) it would be $\kappa \s^2/2$. Now  in 2D for linear elasticity, the complementary energy is
\be\label{ule} U_1(\s_1,\s_2)=\frac{1}{2}K_{ij}\s_i\s_j,  \quad K=C^{-1}.\ee
Thus one might be tempted to replace $\s_i$ by $Z_d(\s_i)$ above and to consider the complementary energy candidate 
\be\label{badr}  U_\ast(\s_1,\s_2)=K_{11}Z_d^2(\s_1)/2+K_{22}Z_d^2(\s_2)/2+K_{12}Z_d(\s_1)Z_d(\s_2).\ee
The problem with this is that the resulting strain-stress relations (partial derivatives of $U_\ast$) are not continuous functions of $\s_i$. While $Z_d^2(\s)$ is continuously differentiable in $\s$,  $Z_d(\s)$  is not. So while the first two terms above are {}{continuously differentiable, the mixed third term involving $K_{12}$ is not, and \eqref{badr} is not satisfactory}. This problem is easily fixed by modifying the third term, {}{and replacing it by the simplest  possible coupling between $\s_1$ and $\s_2$.  We thus  choose the complementary energy density for our constitutive model to be}
\be\label{U} \ur(\s_1,\s_2)=\frac{1}{2}K_{11}Z_d^2(\s_1)+\frac{1}{2}K_{22}Z_d^2(\s_2)+K_{12}\s_1 \s_2 ,\ee
where $d=1/\sqrt{\rho}$.  This is once continuously differentiable (but only piecewise twice).   {}{Its partial derivative with respect to $\s_i$, namely $\e_i$, depends on $\s_i$ (with same index $i$) in a  piecewise linear fashion, with a change of slope  when $\s_i$ changes sign.} Assuming $C$ and hence $K$ to be positive definite, one can show that $\ur$ is  strictly convex. Thus the strain-stress relations are invertible and piecewise linear, and so are the stress-strain relations, while the associated strain energy is strictly convex, piecewise quadratic, and once continuously differentiable.  Also, \eqref{U} coincides with the linear elastic  complementary energy 
 \eqref{ule}  whenever both $\s_i\ge0$ (in the first quadrant of the principal stress plane). The stiffnesses change though whenever one or both of the $\s_i$ become negative.  Thus $\ur $ coincides with a different quadratic function within each of the four quadrants of the principal stress plane.
In particular, noting that {}{$K=C^{-1}$, or}
 \[ K=[K_{ij}]=
\frac{1}{\ga^2-\gb^2} \begin{pmatrix} 
      \ga &- \gb \\
    - \gb & \ga\\
   \end{pmatrix},
\]
 one can write 
 \be\label{UU}\ur (\s_1,\s_2)=\frac{1}{2}\hat K_{ij}\s_i\s_j, \ee
 where the matrix $\hat K$, apart from $\rho$, also depends on $\s_i$ in a piecewise constant fashion. Specifically, it depends only on the signs of $\s_i$,  and takes the following four values in the corresponding four quadrants of the principal stress plane (ordered counterclockwise).
 \be\label{four} \begin{pmatrix} 
      K_{11} &K_{12} \\
    K_{12} & K_{22} \\
   \end{pmatrix},\quad \begin{pmatrix} 
      K_{11} /\rho&K_{12} \\
    K_{12} & K_{22} \\
   \end{pmatrix},\quad \begin{pmatrix} 
      K_{11}/\rho &K_{12} \\
    K_{12} & K_{22}/\rho \\
   \end{pmatrix},\quad \begin{pmatrix} 
      K_{11} &K_{12} \\
    K_{12} & K_{22}/\rho\\
   \end{pmatrix}. \ee
   {}{The strain-stress relations are easily obtained  from \eqref{U} using \eqref{ues},  noting that $K_{11}=K_{22}$:
   \be\label{esrel}\e_i=K_{11}\zrm(\s_i)+K_{12}\s_j,\quad j\not=i.\ee
  }
  It is possible to construct the strain energy  density out of  {}{ \eqref{U}, \eqref{four}. For a quadratic complementary energy of the form \eqref{UU}, the corresponding strain energy is also quadratic:}
  \be\label{WW}\wwr (\e_1,\e_2)=\frac{1}{2}\hat C_{ij}\e_i\e_j, \ee
  where  $\hat C=\hat K^{-1}$.  {}{Since $\ur$ is piecewise quadratic,  so is $\wwr$. } In particular, $\hat C$  is a piecewise constant matrix that takes four values, the inverses of \eqref{four}, in four sectors of the principal strain plane that are the images of the four quadrants of the stress plane under the mapping $(\s_1,\s_2)\mapsto (\e_1,\e_2)$  {}{defined by the strain-stress relations \eqref{esrel}.} See Fig.~\ref{fig:subfigure1}. {}{Switching between these four sectors occurs at points on the uniaxial stress lines; these straight lines are where the principal stresses change signs. See 
  Fig.~\ref{fig:subfigure2}.}  Thus for example in the sector corresponding to $\s_1>0$, $\s_2<0$, the value of $\hat C$ in \eqref{WW} {}{is equal to the inverse of the fourth matrix in \eqref{four}:}
  \be\label{CC}\hat C=
  \frac{(\ga^2-\gb^2)\rho}{\ga^2-\gb^2 \rho} \begin{pmatrix} 
\ga/\rho& \gb  \\
 \gb  & \ga
 \end{pmatrix},  \quad\text{provided  }\;  \ga\e_1/\rho+\gb\e_2>0, \quad \gb\e_1+\ga\e_2<0.
\ee
The two inequalities above define the sector in the principal strain plane that corresponds to the quadrant $\s_1>0$, $\s_2<0$.  Whenever these two inequalities hold,  the stress-strain law {}{ is arrived at  by differentiating \eqref{WW}  (with $\hat C$ as in \eqref{CC}) with respect to $\e_i$. The result is}
\be\label{se} \s_1=h(\rho)(\ga\e_1/\rho+\gb\e_2), \quad \s_2= h(\rho)(\gb\e_1+\ga\e_2),\quad\text{where  }\; h(\rho)= \frac{(\ga^2-\gb^2)\rho}{\ga^2-\gb^2 \rho}.\ee

\subsection{3D Constitutive Law}\label{SS41}
In 3D (letting Latin indices range over $\{1,2,3\}$), the linear elastic isotropic strain energy function in terms of principal strains  is 
\[ W(\e_1,\e_2,\e_3)=C_{ij}\e_i\e_j, \qquad C=
\begin{pmatrix}
 \ga & \gb & \gb \\
 \gb & \ga & \gb \\
 \gb & \gb & \ga
\end{pmatrix},
\]
where  $\ga=2\mu+\lambda$, $ \gb=\lambda$ as before. Positive definiteness of $C$ is equivalent to 
\be\label{pd3}\ga-\gb>0,  \quad \ga+2\gb>0\ee
(equivalent to  $\mu>0$, $\;3\lambda+2\mu>0$). The (principal) stress-strain relations are
\[\s_i=C_{ij}\e_j.\]
For the compression-weakening material we construct the complementary energy in analogy to \eqref{U}
\[ U(\s_1,\s_2,\s_3)   =\      \frac{1}{2}\sum_{i=1}^3K_{ii}  Z_d^2(\s_i)+    \frac{1}{2}\sum_{i=1}^3\sum_{\;\; j=1, j\neq i}^3  K_{ij}\s_i\s_j   ,   \]
where
\[ K=\frac{1}{(\ga-\gb) (\ga+2 \gb)}\begin{pmatrix} \alpha +\beta  & -\beta  & -\beta  \\
 -\beta  &\alpha +\beta& -\beta  \\
 -\beta  & -\beta  & \alpha +\beta
\end{pmatrix}. \]
In particular,  this function is continuously differentiable and piecewise quadratic. The second derivatives are piecewise constant and suffer jump discontinuities across the planes $\s_i=0$ in 3D principal stress space.  Thus $U$ equals a quadratic function in each octant. In particular, in the octant 
\be\label{oct}\s_1>0, \quad \s_2<0, \quad \s_3<0, \ee
we have that 
\[U(\s_1,\s_2,\s_3) =\hat  K_{ij}\s_i\s_j   , \qquad \hat K=\frac{1}{(\ga-\gb) (\ga+2 \gb)}\begin{pmatrix} \alpha +\beta  & -\beta  & -\beta  \\
 -\beta  & \frac{\alpha +\beta }{ \rho} & -\beta  \\
 -\beta  & -\beta  & \frac{\alpha +\beta }{ \rho}
\end{pmatrix} . \]
In this octant,  the strain-stress relations read $\e_i=\hat K_{ij}\s_j$.  These are  invertible for $0<\rho\le1$. The inverse can be found explicitly.  The stress-strain relations $\s_i=\hat C_{ij}\e_j$ with $\hat C=\hat K^{-1}$ ( $\hat K$ as above) are valid in the sector of principal strain space  which is the image of the octant \eqref{oct}.  

\subsection{Mechanical Behavior}\label{C4}
 
The stress-strain relations due to the constitutive law just constructed are as follows. Let  
$\hat\s_i(\e_1,\e_2)=\p \wwr (\e_1,\e_2)/\p\e_i$,
with $\wwr$ the strain energy density in terms of principal strains from \eqref{WW}.  Suppose the spectral representation of the strain tensor  is\footnote{Here for example $\bm\otimes\bu$  is the tensor with components $m_i u_j$.}  $\bE=\sum_{i=1}^2\e_i \bv_i\otimes\bv_i$, with $\e_i$ the eigenvalues (principal strains) and $\bv_i$ the eigenvectors of $\bE$.
 Define the tensor function  
 $$\hat \bS(\bE)=\sum_{i=1}^2\hat\s_i(\e_1,\e_2) \bv_i\otimes\bv_i.$$  
 The tensor stress-strain relation is $\bS=\hat \bS(\bE)$, so that the principal stresses are given by $\hat\s_i(\e_1,\e_2)$  in terms of the principal strains, while $\bS$ has the same eigenvectors $\bv_i$ as $\bE$.  The inverse, strain-stress relation, is given by $\bE=\hat\bE(\bS)$, where
 $$\hat \bE(\bS)=\sum_{i=1}^2 \hat\e_i(\s_1,\s_2) \bv_i\otimes\bv_i,\quad \bS=\sum_{i=1}^2\s_i \bv_i\otimes\bv_i, \quad \hat\e_i(\s_1,\s_2)=\frac{\p U(\s_1,\s_2) }{\p\s_i}.$$
The generalization of the above to 3D is immediate and we omit it.
 
Define the elastic constants
\[ E=\frac{(\ga-\gb) (\ga+2 \gb)}{\ga+\gb}, \qquad \nu=\frac{\gb}{(\ga+\gb)}.\] 
For a linear elastic isotropic solid these are  Young's Modulus and Poisson's Ratio, respectively. For our 3D model, we consider uniaxial stress with principal stresses $\s_i$  ($\s_2=\s_3=0$) and corresponding principal strains $\e_i$. We find
 that 
 \be\label{us}\s_1 =\begin{cases} 
E\e_1,&  \e_1\ge 0,
\\  
\rho E\e_1,&\e_1<0,\end{cases}
\qquad 
\e_2=\e_3  =\begin{cases} 
-\nu\e_1,&  \e_1\ge 0,
\\  
-\rho \nu\e_1,&\e_1<0.\end{cases}\ee
In other words, there is a loss of stiffness  in compression: the effective Young's Modulus (defined as the ratio of longitudinal stress and strain  in uniaxial stress)  is $E$ for tension and $\rho E$ for compression. The uniaxial stress-strain relation thus has the form of Fig.~\ref{subfig1b}.  At the same time, there is a weakening of the Poisson effect: the effective  Poisson's Ratio---defined as minus the ratio of transverse to longitudinal strain in uniaxial stress---is equal to $\rho\nu$ in compression, less than the usual value of $\nu$ in tension. {}{This weakening of the Poisson effect in compression is confirmed by simulations of a discrete fiber network model described in Section  \ref{disc}.}  We note that \eqref{us} can be written as $\s_1=\zr(E\e_1)$, $\e_2=\e_3=-\zr(\nu\e_1)$ in view of \eqref{zr}. The bulk modulus also decreases by a factor of $\rho$ from hydrostatic tension to compression. 

We turn to simple shear; a 2D description is sufficient.  The strain tensor in simple shear has component matrix
\[  [\bE]= \begin{pmatrix} 
      0 &\gamma/2 \\
    \gamma/2 & 0\\
   \end{pmatrix},
\]
where $\gamma$ is called the amount of shear.  One notes that the principal strains change signs together with $\gamma$, (provided one maintains the order of the eigenvectors of $\bE$). 
It is easy to verify that for $\gamma>0$ and  $\gamma<0$ the principal strains lie in different sectors of the principal strain plane (where different quadratic branches of the energy  function are in force).  Accounting for this, one can compute the matrix of components of the stress tensor (in the same basis). The result is
\be\label{nsss}  [\bS]= \begin{pmatrix} 
      N(\rho)|\gamma| &M(\rho)\gamma \\
    M(\rho)\gamma &  N(\rho)|\gamma|\\
   \end{pmatrix}  ,
\ee
where
\[      M(\rho)= \frac{(\ga^2-\gb^2) [\ga- \gb\rho+(\ga- \gb)\rho]}{4(\ga^2-\gb^2\rho)} ,\quad    
N(\rho)= \frac{\ga(\ga^2-\gb^2) (1-\rho)}{4(\ga^2-\gb^2\rho)}. \]
Here $M(\rho)$ is an effective shear modulus (ratio of shear stress and amount of shear for simple shear).  It is positive for any $0<\rho\le1$. For $\rho=1$ it equals the usual shear modulus $\mu$. We observe that the shear stress-strain relation is linear (with no discontinuities in slope), but the slope depends on the stiffness ratio $\rho$. {}{The shear modulus  $M(\rho)$  is an increasing function of  $\rho$, from about $0.8\mu$ at $\rho=0$ to the linear elastic value $\mu$  at $\rho=1$. This is plausible, since compression weakening (lower $\rho$)  also causes a decrease of shear stiffness.}

At first glance,  it seems surprising that there are normal stresses present in simple shear under small strain conditions. These occur in \eqref{nsss} provided there is compression weakening ($\rho<1$). For $\rho=1$,   the normal stresses vanish, $N(1)=0$, as expected for isotropic linear elasticity.  The normal stress modulus $N(\rho)>0$ for $\rho<1$ is the ratio of normal stress to the magnitude $|\gamma|$ of the amount of shear.  {}{The normal stress modulus is a decreasing function of $\rho$, and vanishes at $\rho=1$.}  For $\rho<1$, normal stresses are positive regardless  of the sign of $\gamma\not=0$; thus to maintain the simple shear, hydrostatic tension must be applied in addition to the shear stress; this hydrostatic tension is often called ``negative normal stress'' in the literature  \cite{11janmey,conti}, since it corresponds to  negative pressure.  It corresponds to a reverse or negative Poynting effect  \cite{poynting,mihai}.  This phenomenon has been observed and studied as a somewhat unusual characteristic of fibrous hydrogels (networks of semiflexible biopolymers) \cite{11janmey,conti,notbth}. It is understood \cite{conti} that the underlying mechanism is loss of compression strength of fibers in the direction corresponding to the compressive principal strain in simple shear. This corresponds to compression weakening, and thus explains why our model is capable of predicting the reverse Poynting effect. For more details, see the Discussion  (Section \ref{disc}.)

\section{The Contracting Cell Problem in 2D}\label{sec3}

\subsection{Formulation and Solution}\label{SS1}
 We model the situation of a contracting cell in a fibrin matrix.  It turns out that radially symmetric solutions can be constructed analytically, so the cell is modelled as a disk  of radius $a$ (in 2D) centered at the origin, while the matrix is the annulus $a<r<A$, where $A$ is the outside radius and $r =|\bx|$ is radial distance from the center, while $\bx$ is the position vector. Displacement fields with radial symmetry are of the form $\bu (\bx)=u( r )\bx/r$ in terms of the radial displacement (scalar) function $u( r)$.   The principal strains and stresses  are functions of $r$:
 \be\label{eu}\e_1=\er( r )=u'( r),\quad  \e_2=\et( r )=u(r  )/r,\ee
 where  a prime indicates a derivative, while $\e_1$ is the radial  strain and $\e_2$ is the circumferential  strain.  The equilibrium equations in terms of the radial stress $\s_1=\sr(  r)$ and hoop stress $\s_2=\st(r )$ reduce to 
 \be\label{eq}(r\sr(r  ))'=\st (r )\ee
 We suppose that  (i) the cell shrinks, and (ii) that the outside boundary of the matrix is traction free. We model (i) and (ii)  by the   boundary conditions
 \be\label{bc1}  u(a)=-u_0,\ee
  where $u_0$ is a \emph{positive} constant, and 
 \be\label{bc2} \sr(A )=0,\ee
  respectively.  The solution of the corresponding  \emph{linear elastic} problem (with $\rho=1$) has the property  that  $\sr (r )>0$, $\st (r )<0$ for $a<r<A$. Adopting these inequalities \emph{a priori}   as an \emph{ansatz} in the case of the compression weakening material  ($0<\rho<1$), the stress-strain relations are given by \eqref{se}; hence the second boundary condition \eqref{bc2} becomes
  \be\label{bc3}   \ga u'(A)/\rho+\gb u(A)/A=0.\ee
 Substituting \eqref{eu} into \eqref{se}, and the result into \eqref{eq}, yields a 2nd order linear ODE for $u(  r )$:
 \be\label{ode} r^2u''( r )+ru'(  r  )-\rho u(r)  =0 \quad\text{for}\;\;a<r<A;\ee
  $u(r)$ is also subject to the boundary conditions \eqref{bc1} and \eqref{bc3}. The solution of this boundary value problem is admissible provided it can be verified \emph{a posteriori} that it satisfies the \emph{ansatz} 
 \be\label{ass} \sr (r )>0,\quad \st (r )<0\quad\text{for}\;\;a<r<A\ee
which ensures that \eqref{se} holds.

The general solution of the  ODE \eqref{ode}  for $u(  r )$ is (letting $\xi=\sqrt{\rho}$)
\be\label{ur}  u( r )=c_1 r^{-\xi} + c_2 r^{\xi},   \qquad   \xi=\sqrt{\rho}\ee
The constants $c_1$ and $c_2$ are obtained by satisfaction of the boundary conditions \eqref{bc1} and \eqref{bc3}. 
The result is 
\be\label{const}c_1=
-u_0\frac{ a^{\xi}A^{2 \xi} \left(\alpha +\beta  \xi\right)}{a^{2 \xi}
   \left(\alpha -\beta  \xi\right)+A^{2 \xi} \left(\alpha +\beta  \xi\right)}
 , \quad   
 c_2= -u_0 \frac{a^{\xi} \left(\alpha -\beta  \xi\right)}{a^{2 \xi}
   \left(\alpha -\beta  \xi\right)+A^{2 \xi} \left(\alpha +\beta  \xi\right)}
   \ee
The radial displacement \eqref{ur} then takes the form
\be\label{urc} u(r)= -u_0\frac{ \left(\alpha +\beta  \xi\right) \left(\frac{r}{A}\right)^{ -\xi}+  \left(\alpha-\beta  \xi \right) \left(\frac{r}{A}\right)^{ \xi}}{\left(\alpha +\beta  \xi\right) \left(\frac{a}{A}\right)^{ -\xi}+  \left(\alpha-\beta  \xi \right) \left(\frac{a}{A}\right)^{ \xi}}  ,   \qquad a\le r\le A.
\ee
The stresses are given by 
\begin{align} \label{sss1}
 \sr (r)=& (u_0/a) \frac{\xi  (\alpha^2 -\beta^2 ) 
   \left[  \left(\frac{r}{A}\right)^{-\xi-1}-  \left(\frac{r}{A}\right)^{\xi-1}\right]}{\left(\alpha +\beta  \xi\right) \left(\frac{a}{A}\right)^{ -\xi-1}+  \left(\alpha-\beta  \xi \right) \left(\frac{a}{A}\right)^{ \xi-1}}  ,
 \\ \label{sss2}
\st (r)=&-(u_0/a)\frac{\xi^2  (\alpha^2 -\beta^2 ) 
   \left[  \left(\frac{r}{A}\right)^{-\xi-1}+ \left(\frac{r}{A}\right)^{\xi-1}\right]}{\left(\alpha +\beta  \xi\right) \left(\frac{a}{A}\right)^{ -\xi-1}+  \left(\alpha-\beta  \xi \right) \left(\frac{a}{A}\right)^{ \xi-1}}   , \qquad  a\le r\le A.
\end{align}
In order to verify the ansatz  \eqref{ass}, we rewrite them as 
\begin{align}\label{str}
 \sr (r)=& u_0 \frac{\xi a^{\xi} (\alpha^2 -\beta^2 )  r^{-\xi-1}
   \left(A^{2 \xi}-r^{2 \xi}\right)}{a^{2 \xi} \left(\alpha -\beta  \xi\right)+A^{2 \xi} \left(\alpha +\beta  \xi\right)}  ,
 \\\label{str2} \st (r)=&-u_0
 \frac{\xi^2 a^{\xi} (\alpha^2 -\beta^2 )  r^{-\xi-1} \left(A^{2
   \xi}+r^{2 \xi}\right)}{a^{2 \xi} \left(\alpha -\beta  \xi\right)+A^{2
   \xi} \left(\alpha +\beta  \xi\right)}  , \qquad  a\le r\le A.
\end{align}
In view of \eqref{pd}  and since $0<\rho\le 1$   ($0<\xi\le 1$) we have $\ga\pm\gb\xi>0$. Also $A^{2 \xi}\pm r^{2 \xi}\ge 0$  and  $u_0>0$. It follows from the above form that the inequalities \eqref{ass} are satisfied. Therefore,  \emph{\eqref{urc} with stresses
\eqref{sss1}, \eqref{sss2} provides the solution to the contracting cell problem for the constitutive law provided by the compression weakening model of Section \ref{sec2}.}

The solution is proportional to the contractile displacement $u_0$ for  $u_0>0$ (contraction). For  $u_0<0$ (expanding cell) the ansatz  \eqref{ass} is violated. The solution for this case is obtained in Section \ref{SS4}; its implications are discussed in Section \ref{disc}.

In the special case $\rho=1$ ($\xi=1$)  we recover the linear elastic solution. The displacement is of the form $u(r)=c_1/r+c_2r$, while stresses are of the form $\s(r)=c_3/r^2+c_4$.

\subsection{Universal Bounds}\label{SS2}  A glance at the general solution \eqref{ur} shows that it contains a term that decays as $r$ increases but also one that increases. That raises the question whether the second term would dominate for large $r$.   Recall that $a\le r \le A$. Because of the boundary conditions, the constants $c_1$ and $c_2$ depend on $A$; see \eqref{const}. It turns out that $c_2$ decreases as $A$ increases and actually vanishes in the limit of an infinite matrix (as $A\to\infty$). In contrast, $c_1$ does not vanish in this limit.  In fact, using the inequalities \eqref{pd}, $0<\xi\le 1$, we have  $\ga\pm\gb\xi>0$ so that rewriting \eqref{urc},
\begin{align*} u(r)=& -u_0\frac{ A^{2 \xi} \left(\alpha +\beta  \xi\right)(r/a)^{ -\xi}+ a^{2\xi}  \left(\alpha-\beta  \xi \right)(r/a)^{ \xi}}{a^{2 \xi}
   \left(\alpha -\beta  \xi\right)+A^{2 \xi} \left(\alpha +\beta  \xi\right)} ,    \quad a\le r\le A \quad\Rightarrow
  \\  |u(r)|\le& u_0 \frac{ A^{2 \xi} \left(\alpha +\beta  \xi\right)(r/a)^{ -\xi}+ a^{2\xi}  \left(\alpha-\beta  \xi \right)(r/a)^{ \xi}}{A^{2 \xi} \left(\alpha +\beta  \xi\right)}=
u_0(r/a)^{ -\xi}+u_0 \frac{ a^{2\xi}  \left(\alpha-\beta  \xi \right)(r/a)^{ \xi}}{A^{2 \xi} \left(\alpha +\beta  \xi\right)}
\end{align*}
 But since  $0< r \le A$  we have  $ (a/A)^{2\xi}\le  (a/r)^{2\xi}$, so that the above gives the following upper bound for the displacements: 
  \[ |u(r)| \le  u_0M_1(r/a)^{- \xi}, \quad M_1= \frac{ 2\ga}{ \left(\alpha +\beta  \xi\right)}
 \]
Since  $u_0>0$, $\ga\pm\gb\xi>0$,  \eqref{ur}, \eqref{const} imply that each of the terms in \eqref{ur} is negative.  Also 
$-c_1>2u_0 a^\xi$ since $a<A$. This leads to the lower bound
\[|u(r)|=-u(r)\ge 2u_0(r/a)^{-\xi}\]
Combining the last two bounds we have the following bound for the displacements induced by a circular cell of radius $a$ contracting radially with displacement $u_0$ in a circular matrix of arbitrary radius.
\be\label{ubound} u_0M^-_1(r/a)^{- \xi}\le |u(r)| \le  u_0M^+_1(r/a)^{- \xi}, \quad M^+_1= \frac{ 2\ga}{ \left(\alpha +\beta  \xi\right)},\quad M^-_1=2
 \ee
The bound is \emph{universal} in the sense that it is \emph{independent of the outside radius} $A$ and shows that the displacements decay with order $O(r^{-\sqrt{\rho}})$, despite the presence of the second (growing) term in \eqref{ur}.  The lower bound in \eqref{ubound} ensures that in fact $u(r)$ does not decay \emph{faster}  than  $r^{-\sqrt{\rho}}$ (which is not guaranteed by the upper bound alone).  This ensures that the decay is slower than the linear elastic one.
 
 A similar calculation based on \eqref{str}, \eqref{str2}  (noting for example that $A^{2 \xi}+r^{2 \xi}\le 2 A^{2 \xi}$ ) gives  universal bounds for the norm of the stress tensor $\s(r)=\sqrt{\sr^2(r)+\st^2(r)}$ in the form
 \be\label{sbound}  (u_0/a)M^-_2 \left(\frac{r}{a}\right)^{-(\sqrt{\rho}+1)}\!\! \le  |\sigma(r)|\le (u_0/a)M^+_2 \left(\frac{r}{a}\right)^{-(\sqrt{\rho}+1)} 
  , \quad M^+_2= \frac{2\xi (\alpha^2 -\beta^2 ) }{\alpha +\beta  \xi},\; M^-_2=\xi M^+_2/4.
 \ee
 where  the constants
 $M^\pm_2$
  depends only on  material properties $\ga$,  $\gb$  and $\rho=\xi^2$ but not on $a$, $A$.
 We conclude that  \emph{stresses and displacements induced by a contracting cell in a matrix composed of  compression weakening material, decay slower than in a linear elastic matrix}  where $u=O(r^{-1})$  and $\s=O(r^{-2})$.
\subsection{Infinite Matrix}\label{SS7} Taking the limit as 
$A\to\infty$ we obtain the displacement due to  a contracting cell in an infinite matrix (with the stress approaching zero at large distances)
\[u(r)=-u_0\left(\frac{r}{a}\right)^{-\xi},  \quad  \xi=\sqrt{\rho}.\]
The stresses are
\[\sr(r)=(u_0/a) \frac{\xi (\alpha^2 -\beta^2 ) }{\alpha +\beta  \xi}  (r/a)^{-\xi-1},  \qquad \st(r)=-\xi\sr(r), \quad a\le r<\infty\]
Thus for a compression weakening material with $\rho<1$  ($\xi<1$)   the displacements, $u(r)=O(r^{-\xi})$, and the stresses, $\s(r)=O(r^{-\xi-1})$ as $r\to\infty$.  Thus both  decay slower than their linear elastic counterparts,  which are $O(r^{-1})$  and $O(r^{-2})$, respectively.  The lower the compression stiffness ratio $\rho=\xi^2$, the slower the decay. We conclude that  \emph{regardless of whether cell mechanosensing is based on stresses or displacements, if cells sense each other mechanically, they can do so} over larger distances in fibrin networks than in materials that do not weaken in compression.

\subsection{The Case of Zero Compression Strength}\label{SS3}

The case $\rho=0$ is interesting but tricky. The limit as $\rho\to 0$  in the solution \eqref{urc}, \eqref{sss1}, \eqref{sss2} is 
\be\label{triv}u(r)=-u_0,  \quad \sr(r)=\st(r)=0, \quad a\le r\le A.\ee 
Taking the limit as $\rho=0$ in the constitutive law  \eqref{se},  yields
\[\s_1=\frac{\alpha^2 -\beta^2}{\ga}\e_1,  \quad  \s_2=0\]
where $\s_1=\sr$,  $\e_1=u'(r)$ for the radially symmetric problem.
Then the equilibrium equation \eqref{eq}  becomes $(r\sr(r  ))'=0$ or $(ru'(r))'=0$. The general solution is 
\[  u(r)=c_1\log r+c_2,     \quad  \sr(r)= \frac{\alpha^2 -\beta^2}{\ga}\frac{c_1}{r}\]
If we enforce the boundary conditions \eqref{bc1}, \eqref{bc2}, the second demands $c_1=0$ and the first that $c_2=u_0$, thus for the bounded traction free matrix we recover \eqref{triv}. For the infinite matrix however, since the stress decays as $1/r$, we only have \eqref{bc1} to enforce and that leaves a one parameter family of solutions
\[ u(r)=c_1\log(r/a)-u_0\]
If we insist though that $u(r)$ remain bounded, then necessarily $c_1=0$ and  the only solution is \eqref{triv}.

\subsection{Expanding cells are short sighted}\label{SS4}
We ask the following question:
\emph{Suppose the cell has a choice between contracting and expanding. What is more efficient for mechanosensing in a fibrin network matrix?}

To answer this in the context of our model, suppose now that we change the sign in \eqref{bc1} and require $u_0<0$.  Then the signs in \eqref{sss1} and \eqref{sss2} are reversed, and \eqref{ass} is violated! The previous solution with a mere sign change does not apply here. One expects that changing the sign of $u_0$  will reverse the signs in \eqref{ass},  so that instead of \eqref{ass}, we will make the ansatz
 \be\label{mass} \sr (r )<0,\quad \st (r )>0\ee
This will put the stresses in the second, as opposed to the fourth quadrant of the principal stress plane, involving a different quadratic branch of the energy function. Then $\hat C$ in \eqref{WW} will equal the inverse of the second, instead of the fourth matrix in \eqref{four}. Accordingly \eqref{se} must be replaced by
\be\label{mse} \s_1=h(\rho)(\ga\e_1+\gb\e_2), \quad \s_2= h(\rho)(\gb\e_1+\ga\e_2/\rho),\ee
with \eqref{eu} still in force.  This eventually results in a different ODE for $u(r)$, namely
\be\label{mode} r^2u''( r )+ru'(  r  )-\rho^{-1} u(r)  =0 \quad\text{for}\;\;a<r<A.\ee
The general solution is
\be\label{mur}  u( r )=c_1 r^{- d} + c_2 r^{ d},   \qquad    d=1/\sqrt{\rho}\ee
The main difference here is that always $ d>1$ and $ d\to\infty$ as $\rho\to0$. For example the solution of the cell in the infinite matrix is
\[u(r)=-u_0\left(\frac{r}{a}\right)^{- d},  \quad   d=1/\sqrt{\rho}.\]
while the stresses behave as $\s(r)=O(r^{- d-1})$. Thus, both  displacements and stresses induced by an expanding cell decay \emph{faster}  than their linear elastic counterparts,  which are $O(r^{-1})$  and $O(r^{-2})$, respectively.  The lower the compression stiffness ratio $\rho=1/ d^2$, the \emph{faster}  the decay. 

This result agrees with a numerical calculation in \cite{notbth}:  when a concentrated load is exerted in a finite element network model, the displacements decay faster (than linear elasticity predicts) in the direction towards which the force points (the ``pushing'' direction), and slower in the opposite, or pulling direction. 

Thus cell expansion is not a good mechanism for long range mechanosensing,  because displacements due to an expanding cell decay rapidly over space. In a linear elastic matrix, cell expansion and contraction produce fields with the same decay rate.

\section{The Contracting Spherical Cell in 3D}\label{sec4}

In 3D the cell is modelled as a sphere  of radius $a$ centered at the origin, while the matrix is the region $a<r<A$, or the portion of the sphere of radius $A$ outside the cell. Again, $r =|\bx|$ is radial distance from the center, while $\bx$ is the position vector. Displacement fields with radial symmetry are of the form $\bu (\bx)=u( r )\bx/r$ as in 2D.   The principal strains and stresses  are functions of $r$:
 \be\label{eu3}\e_1=\er( r )=u'( r),\quad  \e_2=\e_3=\et( r )=\ef(r)=u(r  )/r.\ee
where $r$,  $\theta$, $\varphi$ are spherical polar coordinates.  Since $\st(r)=\sfi(r)$, the equilibrium equations reduce to 
 \be\label{eq3}(r^2\sr(r  ))'=2r\st (r ),\ee
 We impose  the same  boundary conditions \eqref{bc1}, \eqref{bc2}
 as in 2D.  The solution of the corresponding  \emph{linear elastic} problem ($\rho=1$) has the property  that  $\sr (r )>0$, $\st (r )=\sfi(r)<0$ for $a<r<A$. Once again we presuppose \eqref{ass},  to be verified later.
 Substituting \eqref{eu3} into the  3D stress-strain relations, and the result into \eqref{eq3}, yields a 2nd order linear ODE for $u(  r )$:
 \be\label{ode3}  r^2 u''(r)+2r  u'(r)-\frac{2 \ga  \rho}{\ga+\gb(1-  \rho)} u(r)=0     \quad\text{for}\;\;a<r<A;\ee
  $u(r)$ is also subject to the boundary conditions \eqref{bc1} and \eqref{bc2}. The solution of this boundary value problem is admissible provided it satisfies 
 \eqref{ass}.  Note that \eqref{ode3} involves the material constants $\ga$ and $\gb$,  in contrast to the ODE \eqref{ode} for 2D.  Letting
 \[g(\rho)=\frac{ \ga  \rho}{\ga+\gb(1-  \rho)} \]
 the general solution of \eqref{ode3} is 
 \be\label{ur3}  u( r )=c_1 r^{\xm} + c_2 r^{\xp},   \qquad   \xi_\pm=\frac{1}{2}\left(-1 \pm\sqrt{1+8 g(\rho
   )}\right) \ee
 Observing that $g(0)=0$, $g(1)=1$ and that $g(\rho)$ is monotone increasing (as can be shown using  \eqref{pd3}), we have that 
 \be\label{ran}-2\le \xm<-1, \quad 0<\xp \le 1.\ee
 Also, $\xm\to  -1$ as $\rho\to 0$. 
 Proceeding as in 2D we find the constants $c_1$ and $c_2$ in \eqref{ur3} from the boundary conditions \eqref{bc1} and \eqref{bc2}. The solution can be written as follows. Define the constants
 \[P_\pm=\pm\left([\ga+\gb(1-\rho)]\xi_\pm+2\gb\rho\right), \quad Q_\pm=\ga+\gb(1+\xi_\pm), \quad R=\frac{(\ga-\gb)(\ga+2\gb)}{(\ga+\gb)^2-\gb(\ga+3\gb)\rho}.
\]
The displacement is 
 \be\label{urc3} u(r)=-u_0\frac{ P_+ \left(\frac{r}{A}\right)^{\xm}+P_- \left(\frac{r}{A}\right)^{\xp}}{P_+ \left(\frac{a}{A}\right)^{\xm}+P_- \left(\frac{a}{A}\right)^{\xp}}
 \ee
 The stresses are
 \begin{align} \label{s31}
 \sr (r)=& (u_0/a) \frac{R P_+P_-
   \left[  \left(\frac{r}{A}\right)^{\xm-1}-  \left(\frac{r}{A}\right)^{\xp-1}\right]}{P_+ \left(\frac{a}{A}\right)^{\xm-1}+P_- \left(\frac{a}{A}\right)^{\xp-1}}  ,
 \\ \label{s32}
\st (r)=&-(u_0/a) \frac{\rho R 
   \left[  Q_-P_+\left(\frac{r}{A}\right)^{\xm-1}+Q_+P_- \left(\frac{r}{A}\right)^{\xp-1}\right]}{P_+ \left(\frac{a}{A}\right)^{\xm-1}+P_- \left(\frac{a}{A}\right)^{\xp-1}}   , \qquad  a\le r\le A.
\end{align}
Using \eqref{pd3},  \eqref{ur3} and \eqref{ran}, one can show that 
\[ P_\pm>0, \quad Q_\pm>0, \quad R>0.\]
This implies that the inequalities \eqref{ass} are satisfied and the solution is admissible.
It is possible to derive global bounds (as for the 2D solution) of the form
\be\label{3db}\begin{split}u_0 M^-_1 \left(\frac{r}{a}\right)^{\xm}\le & |u(r)|\le u_0 M^+_1 \left(\frac{r}{a}\right)^{\xm}, \\ &\\
  (u_0/a)M^-_2 \left(\frac{r}{a}\right)^{\xm-1} \le & |\sigma(r)|\le (u_0/a)M^+_2 \left(\frac{r}{a}\right)^{\xm-1} 
  \end{split}\ee
where $\xm$ is the negative root in \eqref{ur3}, \eqref{ran}, while $M^\pm_1$ and $M^\pm_2$ are constants that depend only on $\ga$, $\gb$ and $\rho$, that is, on material constants only, but not on the geometry (not on $A$, $a$).
  For $\rho<1$, in view of \eqref{ran} we have $\xm>-2$, hence these bounds imply that the decay is always slower than the linear elastic case $\rho=1$ (for which $\xm=-2$, $\xp=1$). 
  
  In the limiting case $\rho=0$, the 3D constitutive law reduces to 
  \[\s_i =\begin{cases} 
E\e_i,&  \e_i\ge 0,
\\  
0,&\e_i<0,\end{cases}, \qquad   i=1,2,3.
\]
For the contracting inclusion, this means that the only nonzero stress component is $\s_r(r)=E u'(r)$ since the hoop stains are compressive,  and the  equilibrium equation \eqref{eq3} reduces to  $(r^2\s(r))'=0$. The implies that $s(r)=c_1/r^2$, hence $u(r)=c_2/r+c_3$ for constants $c_i$.  The solution to the contraction inclusion in an infinite matrix is $u(r)=-u_0a/r$. This validates a heuristic argument  made by Notbohm \emph{et al. }  (see the discussion involving Eq. (1) in \cite{notb}).

\section{Results}\label{sec1}

To test the hypothesis that displacements in a compression weakening elastic material propagate over a longer range than in a linear elastic one, we develop a \emph{homogeneous continuum model}, but one with a nonlinear elastic constitutive law, in which the principal stresses depend on the principal strains in a special,  piecewise linear fashion. Following Notbohm et al. \cite{notb}, we choose the stiffness in compression to be lower than that in tension. In 1D this  is easy to do; such a stress-strain law that weakens in compression  is shown in Fig.~\ref{subfig1b}. The slope (stiffness) is less in compression that in tension. The ratio of the two is a parameter $\rho$ in the range 
\be\label{-1} 0<\rho\le 1.\ee
We call this parameter the \emph{compression stiffness ratio}. When $\rho=1$ we are back to linear elasticity.

It is rather challenging to construct higher dimensional constitutive models that weaken in compression in an acceptable fashion. The stresses have to be continuous functions of the strains (though not necessarily differentiable), and they must be derivable from a strain energy density function, otherwise the elastic constitutive law is thermodynamically unsound.  This task is the subject of Section \ref{sec2}. The result is a strain energy density function that is a piecewise quadratic function of the principal strains. Its level curves in the principal strain plane are shown in Fig.~\ref{fig:subfigure1}, while those of the underlying linear elastic one are shown in Fig.~\ref{fig:subfigure3}.

Once the 2D constitutive model is constructed, it is more straightforward to generalize it to a 3D constitutive law  in Section \ref{SS41}. 

 The mechanical behavior of the constitutive model is described in Section \ref{C4}. 
A comparison with certain types of unusual experimental and simulated behavior characteristic of fibrin can be found in Section \ref{disc}.

In Section \ref{sec3} we consider a  problem intended to model the contracting cell in a matrix  that exhibits loss of stiffness in compression. The matrix may be finite but  possibly large compared to the cell; we do account for external boundaries. We start with 2D. The elastostatic problem for our constitutive model is tractable in the radially  symmetric case. In Section \ref{SS1}, the cell is modelled as a contracting circle of radius $a$; the matrix as a disk of radius $A$, with the cell at its center.  The case $A>>a$ is typical, although we leave $A$ and $a$ arbitrary ($A>a>0$). The matrix external boundary ($r=A$) is  free of applied forces (traction-free), while the cell boundary ($r=a$) suffers a prescribed negative radial displacement $-u_0$.  This gives two boundary conditions. The matrix is assumed to be composed of the material with constitutive law developed in Section \ref{sec2}.  This is characterized by two elastic constants and the stiffness ratio $\rho$.

We obtain the solution to this problem in Section \ref{SS1}. The displacement field is radial, of the form
 \be\label{0}  u( r )=c_1 r^{-\sqrt{\rho}} + c_2 r^{\sqrt{\rho}}, \quad a<r<A, \ee
 so that the first term in \eqref{0} decays, while the second grows, as the distance  $r$ from the origin increases.  The relevant stress components take the general form
 \be\label{1}\sigma_i(r)=c_3 r^{-(\sqrt{\rho}+1)} + c_4 r^{\sqrt{\rho}-1}, \quad a<r<A,\ee
 where $\s_i$ stands for either  $\sr$ or $\st$.
 The constants $c_1$ through $c_4$ are determined by the boundary conditions.  For the complete closed-form solution see \eqref{urc}--\eqref{sss2}. When we set $\rho=1$ we recover the linear elastic solution $u(r)=c_1/r+c_2 r$.  For $0<\rho<1$,  the decreasing  term does decay slower than the corresponding linear elastic term, but the role of the growing terms seems unclear.  We deal with this in Section \ref{SS2}. We observe that the constants $c_2$, $c_4$ tend to  zero in the limit as $A\to\infty$. This suggests that the growing terms may remain small. We find that this is indeed the case and deduce bounds for the displacement and the norm $\s(r)$ of the stress tensor  in the  form
 \be\label{b} 
 \begin{split}u_0 M^-_1 \left(\frac{r}{a}\right)^{-\sqrt{\rho}}\le & |u(r)|\le u_0 M^+_1 \left(\frac{r}{a}\right)^{-\sqrt{\rho}}, \\ &\\
  (u_0/a)M^-_2 \left(\frac{r}{a}\right)^{-(\sqrt{\rho}+1)} \le & |\sigma(r)|\le (u_0/a)M^+_2 \left(\frac{r}{a}\right)^{-(\sqrt{\rho}+1)} 
  \end{split}\ee
These inequalities involve only the negative exponent $-\sqrt{\rho}$, although  they bound the entire displacement and stress in \eqref{0}, \eqref{1}, including the growing terms. Also, they are \emph{universal},  in the sense that the constants $M^\pm_1$ and $M^\pm_2$ are \emph{independent} of the size of the matrix $A$ and the inclusion radius $a$ (the geometry); see \eqref{ubound},  \eqref{sbound}. Rather, they only depend on the elastic constants and the stiffness ratio $\rho$. Also the dependence of the bounds on  $r$ is through $r/a$, or distance measured in cell radii.  

The lower bounds in \eqref{b} prove that \emph{the displacement and stress fields exterior to a contracting spherical inclusion in a compression weakening material (governed by the constitutive model developed here)  decay slower that the corresponding linear elastic ones},  which satisfy analogous bounds, obtained by replacing $\sqrt{\rho}$ by $1$ in \eqref{b}

 {}{The continuum solution \eqref{0} is in good quantitative agreement with  numerical simulations   of the fiber discrete network model of \net.   See Section \ref{disc} and Table 1 for details.}

The case of the infinite matrix is briefly dealt with in Section \ref{SS7}.  In Section \ref{SS3} we also consider the case $\rho=0$, which is a singular limit. 

We  consider hypothetical cells that \emph{expand} instead of contracting in Section \ref{SS4}.  Surprisingly, the solution for a compression weakening material  decays \emph{faster} than the linear elastic one. This explains why expansion, as opposed to contraction, of individual cells is not  conducive to long-range mechanosensing. See Section \ref{disc} for the implications of this.

 The 3D contracting cell problem is formulated in Section \ref{sec4}  as that of a sphere of radius $a$, contracting with radial displacement $-u_0$ at the center of a spherical matrix of radius $A$, whose external boundary is traction free. The 3D constitutive law of Section \ref{SS41} is used. The solution  \eqref{urc3}--\eqref{s32} is more complicated than the 2D one, but qualitatively very similar. The solution has  the form 
\[ u( r )=c_1 r^{\xm} + c_2 r^{\xp}, \]
but the exponents $\xi_\pm$ now depend on the elastic moduli, in addition to the compression stiffness ratio $\rho$.  The  exponents  $\xi_\pm $  satisfy
 \[-2\le \xm<-1, \quad 0<\xp \le 1,\]
while universal  bounds of the form \eqref{3db} are  valid. Once again, in 3D the decay is slower than in a linear elastic matrix where the displacement would decay with $\xm=-2$.  Fits of displacement data from the 3D fiber network model simulations of \net to the form $u=Cr^{-n}$ for $\rho=0.1$ give an exponent $n=0.67$ for a high 3D connectivity of 14, and $n=0.82$  for a low connectivity of 3.5, close to the value considered representative of fibrin, compared to the continuum model prediction of $\xi_-=-1.1$. Experiments involving fibroblasts in 3D fibrin \cite{notb} give $n=0.52$. The agreement is not as good as in 2D, but  the qualitative conclusions remain the same.

\section{Discussion}\label{disc}

\subsection{A Continuum Model for Fibrin}\label{D1}

In Section \ref{sec2} we construct a new hyperelastic constitutive model for  compression-weakening materials. We view this constitutive law as a continuum model for a fibrin network, and possibly other fibrous materials. The essential characteristic we wish to capture is loss of stiffness in compression. In a discrete fibrous network,  this happens because individual fibers  buckle under compression. In our  continuum model, it is embodied in  a special constitutive nonlinearity. Specifically, each principal stress is a  piecewise linear function of the principal strains in a way that generalizes the  behavior  of Figure \ref{subfig1b} (a 1D idealization of buckling behavior) to 2D and 3D.

We compare the behavior of the model  under certain homogeneous deformations described in Section \ref{C4}  with  experimental observations for fibrin.

The  uniaxial stress-strain relation  \eqref{us} predicted by our model  is  linear in tension, and neglects the gradual nonlinear strain stiffening characteristic of many polymeric fibrous materials under tension, including fibrin  \cite{piech,10storm,roeder}. In some cases however,  fibrin  exhibits very nearly linear behavior in uniaxial tension up to strains well beyond the scope of our small-strain model  \cite{12brown,jawerth}.

Choosing  piecewise linear relations for our model has certain advantages. First, it  allows us to solve  analytically some  model problems that  provide insight into mechanosensing by contractile cells. Second,  it is  the simplest model that accounts for compression weakening without introducing other nonlinearities. This minimalist approach allows us to isolate and study the effect of compression weakening on the slow decay of elastic fields due to contracting inclusions.  We find such an effect in the absence of stiffening, as discussed below in Section \ref{D2}.

{}{One  prediction of our model constitutive law is a weakening of  the Poisson effect in uniaxial compression. The effective  Poisson's ratio (minus the ratio of transverse to longitudinal strain in uniaxial stress) is equal to $\nu_c=\rho\nu_t$ in compression, compared to  the value of $\nu_t$ in tension; see \eqref{us}. We are not aware of experiments  reporting both   tensile and compressive values of the effective Poisson ratio for fibrin.  We compare this prediction with   simulations of a discrete  fiber network model \cite{notbth,notb},  where each  fiber has a stress-strain curve  as in Fig.~ \ref{subfig1b}, with stiffness ratio $\rho$. For 2D uniaxial stress, simulation results  for the ratio of the compressive and  tensile  values of the  effective Poisson ratio are within 1\% to the fit   $\nu_c/\nu_t=0.74 \rho$ for the values tested ($\rho=$ 0.1, 0.3, 0.5, 0.7).  The fiber network model undergoes weakening of  the Poisson effect in  compression, in  qualitative agreement  with the prediction of our model, $\nu_c/\nu_t= \rho$, in that the ratio is less than one and  increases linearly with $\rho$. }

Fibrous polymer networks exhibit  unusual  behavior  under homogeneous simple shear:  they develop normal stresses corresponding to a negative hydrostatic pressure, with a sign opposite to that of the usual Poynting effect in nonlinear elasticity of rubberlike solids \cite{poynting,mihai}.  This reverse Poynting effect  is often termed ``negative normal stresses.'' \footnote{Presumably this term refers  to the sign of the pressure, rather than that of the normal stress components, which are tensile, hence positive according the the usual sign convention.} The phenomenon was experimentally observed in fibrin \cite{11janmey} and simulated using discrete fiber network models \cite{conti}.  The underlying mechanism was identified in \cite{conti} as ``compressive buckling of the individual filaments''.  Consistently with this, our model predicts the presence of such normal stresses corresponding to a negative pressure,  whenever the compression stiffness ratio  $\rho$ is less than unity, namely, in the presence of compression weakening; see \eqref{nsss}.  Instead, when  $\rho=1$ and the model reduces to linear elasticity,  normal stresses vanish in simple shear.  

The network model of  \cite{conti} predicts a normal stress that is  quadratic  for small strains,  but becomes proportional to the absolute value of the shear stress for large strains. Our piecewise smooth constitutive law predicts that normal stresses in simple shear, and normal strains under pure shear stress, are each proportional to the absolute value of the shear stress.  This result is very similar to Fig.~4 of \cite{conti} (for small bending stiffness) and to the experimental data shown in Fig.~4(b) of \cite{11janmey}.  For pure shear stress, our results agree with the network model simulation results of  \cite{notb}  (supplemental Fig.~S6) in that the ratio of normal strain to  the absolute value of shear strain is constant, negative, and increases in magnitude as the the stiffness ratio $\rho$ decreases from 1 to 0.

Another unusual type of behavior experimentally observed in fibrin \cite{12brown} and exhibited by the network model of \cite{notbth,notb} is termed  ``negative compressibility,'' referring to  a decrease of volume during uniaxial tension. This only occurs at large strains  above 10\% in both experiments and simulations  \cite{12brown,notbth,notb}. This is not a surprise, since such behavior  \emph{in the small strain regime} for an isotropic material would require a negative bulk modulus, which is inconsistent with the positive definiteness of the elasticity tensor in linearized elasticity. Our model behaves as an ordinary linear elastic isotropic solid during uniaxial tension, and  does not exhibit such an effect.  
 
These remarks suggest that negative compressibility is not the primary factor in the observed  slow decay of displacements due to contractile cells in fibrin, as the latter phenomenon does not seem to require large deformations  \cite{notbth}. As we discuss next, the model supports the conclusions of \net  that the primary cause is compression weakening.  

\subsection{Spatial Decay of Elastic Fields Due to Contractile Cells}\label{D2}

 In Section \ref{sec3} we model a contractile cell in a fibrin matrix as a  contracting circular inclusion embedded in a material governed by our model constitutive law in 2D. This is the continuum analog of the 2D contracting inclusion problem for a finite element model of a fiber network  \cite{notbth,notb},  with individual fibers having a force-elongation relation  as in Fig.~\ref{subfig1b} (where the ratio $\rho$ of slopes in compression and tension is chosen between $0$ and $1$).
 The solution \eqref{0}, \eqref{1} to the continuum contracting inclusion problem, together with the bounds \eqref{b}, clearly predicts that both displacements and stresses  decay slower with distance from a contractile inclusion in a material that weakens in compression than in a linear elastic one. 
 
  \net performed simulations of the contracting inclusion using a finite element model of a  fiber network with each node acting as a hinge for the elements  terminating at it, randomness in nodal positions, and different values of  the connectivity  (the average number of fibers meeting at a node).  Each fiber has a stress-strain curve  as in Fig.~ \ref{subfig1b}, with stiffness ratio $\rho$. The numerical  radial displacement data were fit to the form 
\be\label{not}u(r)=C_1 r^{-n}+C_2 r^n,\ee 
with fitting parameters $C_1$, $C_2$, $n$. This form  is consistent with the continuum solution  \eqref{0} of the present work.
 The value  $\rho=0.1$ was used in the simulations \cite{notb}.  The present continuum model also involves $\rho$; the corresponding solution \eqref{0} predicts that $n=\sqrt{\rho}$.  The  fit for $n$ resulting from the network model data depends on the connectivity  (the average number of fibers meeting at a node) of the network simulated.  The network with the highest connectivity of $8$ is likely to behave  closest to the continuum model.  In this case fits of the numerical data give $n=0.36$. Our continuum solution \eqref{0} for the choice  $\rho=0.1$ yields the prediction $n=0.32$. 
  For $\rho=1$ our model reduces to linear elasticity and does not weaken in compression. The discrete network model fit yields $n=0.89$  for $\rho=1$. 

Additional simulations were performed using a network of connectivity 8 and no randomness in nodal positions, for three different values of $\rho$ and for both a contracting and an expanding inclusion, for the purpose of comparing the network and continuum models.  Data were fit to the form \eqref{not} and also to
\be\label{notp}u(r)=C_1 r^{-p_1}+C_2 r^{p_2},\ee 
with fitting parameters $C_1$, $C_2$, $p_1$  and $p_2$.  The continuum model predicts $p_1=p_2=n=\sqrt{\rho}$ for the contracting inclusion, and $p_1=p_2=n=1/\sqrt{\rho}$ for the expanding one; see Section \ref{SS4},  eq. \eqref{mur}. The agreement is quite satisfactory.
 The results of the comparison are summarized in Table 1.

 \begin{table}
  \centering
  \subtable[]{
   \centering
\begin{tabular}{||c||c|c|c|c||}
      $\;\:\rho$ & $\sqrt{\rho}$& $n$&$p_1$&$p_2$\\ \hline \hline
    0.3 & 0.547    & 0.532 & 0.531&0.543   \\
    0.1  & 0.316    & 0.337  & 0.340&0.320  \\
    0.03  & 0.173   & 0.226 &  0.235&0.176    \\
        \end{tabular}
    \label{tab1a}
}
\subtable[]{
   \centering
\begin{tabular}{||c||c|c|c|c||}
    $\;\:\rho$& $1/\sqrt{\rho}$ & $n$&$p_1$&$p_2$\\ \hline \hline
    0.3 & 1.83   & 1.87 & 1.87&1.83 \\
    0.1 & 3.16  & 3.16& 3.17&  3.16 \\
    0.03& 5.77  & 5.17 & 5.17 & 5.17  \\
    \end{tabular}
    \label{tab1b}
}
 \caption {(a) Comparison of   the theoretical prediction  $\sqrt{\rho}$ for the decay power (from the solution \eqref{0} of the continuum model) to the  power $n$ from fits of \eqref{not}, and to $p_1$ and   $p_2$ from fits of \eqref{notp} to simulation data using the discrete fiber model of \cite{notb}, for the contracting  inclusion,  for three values of $\rho$. (b) Analogous comparison for the expanding inclusion; the theoretical prediction is $1/\sqrt{\rho}$.}
\end{table}

 When compression weakening is suppressed  ($\rho=1$ in both models) the decay rate of numerical solutions  is comparable to the linear elastic one, while the continuum problem reduces to the linear elastic one.

According to the present model, a contracting inclusion of radius $a$ in an infinite matrix in 2D with stress approaching zero at infinity induces the displacement field
\be\label{infmat}u(r)=-u_0\left(\frac{r}{a}\right)^{-\sqrt\rho},\quad a<r<\infty\ee
while the stress field is proportional to $r^{-(1+\sqrt\rho)}$. Thus the slow decay rate in the presence of compression weakening ($\rho<1$) has unbounded range. The solution does not approach the 2D linear elastic one, $u(r)=-u_0(r/a)^{-1}$, as $r\to\infty$.
 
 To solve the contracting inclusion problem, previous studies \cite{16shokef,shenoy} have used  different nonlinear models that do not include compression weakening, but exhibit strain stiffening at large tensile strains in different forms. Wang et al. \cite{shenoy} consider   tension-driven alignment of fibers as the mechanism responsible for long-range force transmission in fibrous matrices. In their constitutive model, they  account for this by means of a tension-stiffening term.  This nonlinearity would eventually cause higher stiffness in tension than  compression, provided  the cell stretches the matrix far enough into the nonlinear regime to cause an appreciable difference in tensile and compression stiffness. For fibrin, this  could mean  large  tensile strains  induced by the cell, exceeding the stiffening threshold of about  $10\%$  \cite{roeder,12brown,jawerth},   probably  at a high energetic cost, to facilitate mechanosensing. Moreover, as strains decay with distance from the cell, the difference  between tension and compression stiffness would  become negligible  in a tension-stiffening  model, and  the displacement would approach the linear elastic decay rate.  Thus  the slow decay  of elastic fields would be confined to a ``nonlinear zone''  around  the cell. The larger the critical strain for stiffening,   the smaller the size of such a nonlinear zone.

In contrast, microbuckling involves a sudden drop of stiffness in compression, which for fibrin occurs at a very small strain of about 
$4\times 10^{-4}$ \cite{piech,notb}, two orders of magnitude below  the strains required for tension stiffening.  While both compression weakening and tension stiffening occur in fibrin, and  contracting cells induce both compressive (hoop) and tensile (radial) strains,  compression weakening  is  likely to be the dominant  mechanism, since is is engaged at much lower cell-induced strain levels. In the experiments of  \cite{notbth,notb}  slow displacement decay  was observed at cell-induced strains of $2\%$, below  the level required for appreciable stiffening \cite{roeder,12brown,jawerth} and  less than the  high levels required for prediction of long range transmission of forces  by the model of  \cite{shenoy}.

Our model  treats  the buckling strain  as negligible. Hence,  its  behavior is not smooth at zero strain, and does not approach linear  elasticity for  small strains. As a result, long-range propagation of elastic fields occurs without spatial limit, and  for arbitrarily small cell-applied contractile displacement  $u_0$. 
In practice, the cell contractile strain $u_0/a$ must exceed a  small buckling strain, and the nonlinear zone of slow decay  would be large but finite. In  experiments \cite{notbth,notb}, a  transition to a linear elastic spatial decay rate was not observed within the measurement range of  $100\mu m$, of the order of 10 cell radii.

Other simulations of model networks without compression weakening, but with  gradual stiffening in tension and continuous slope at zero strain, predicted  displacement decay rates close to the  linear elastic one, even for  large  cell strain $u_0/a\approx10\%$  \cite{notbth,notb}. These results and also the experiments of  \cite{25rudnicki},  suggest that  strain stiffening is not the primary factor causing the observed slow  displacement decay.

We conclude that microbuckling in fibrous materials, modelled as compression weakening in the continuum setting, is directly responsible for long-range propagation of elastic fields induced by cell contraction, thus it facilitates mechanosensing  in fibrous biopolymer matrices.

Why do cells contract instead of expanding to facilitate  mechanosensing? Probably a good answer to this question is: because they can!  That is, it may only be possible for the cell to exert tension on the matrix,  because the mechanism is essentially ``winches pulling on ropes'', i.e.,  myosin-II motors pulling on actin filaments. So it may be physically difficult for the cell to generate large forces by pushing on the matrix while it is stationary.  Until recently, pushing forces applied by single cells within a fibrous matrix had not  been observed  \cite{notbles}. Pushing forces only seem to occur during invasive cellular migration into the matrix; even then, pushing may result as a reaction force balancing the contraction of cellular protrusions  \cite{notbles,wolf}. Immobile cells contract during the mechanosensing process, before any growth into the matrix occurs \cite{9winer,notb}.

What could have driven cells to evolve so as to exhibit  almost entirely contractile behavior while stationary? To understand this, we consider a related question in Section \ref{SS4}:
{Suppose a stationary cell has a choice between contracting and expanding. Which  is more efficient for mechanosensing in a fibrin network matrix?}  Because of nonlinearity, specifically the constitutive asymmetry between tension and compression, reversing the sign of the applied boundary displacement $-u_0$ does not simply multiply the solution by $-1$, as would happen in linear elasticity. Instead, if the cell expands and pushes at the matrix (let $u_0<0$) the solution \eqref{mur} is still of the form \eqref{0}, \eqref{1} but with different exponents $\pm1/\sqrt{\rho}$ in place of  $\pm\sqrt{\rho}$.  Bounds similar to  \eqref{b} still hold,  but with negative exponent $-1/\sqrt{\rho}$ that  approaches $-\infty$ as $\rho\to0$. As a result, the displacements and  stresses due to an  {expanding} cell in a compression weakening material  {decay faster} than in a linear elastic material. Thus expanding cells would cloak themselves from other cells;  this is  counterproductive for mechanosensing.   We conclude that  {contractile behavior of individual cells in a fibrous, compression weakening matrix, is far more more efficient for long-range mechanosensing than expansion}. It seems that cells have evolved accordingly.

 \subsection*{Acknowledgments}This work was motivated by experiments performed in collaboration with Dr. Ayelet Lesman and Professor David Tirrell under a grant from the National Science Foundation (Division of Materials Research No. 0520565) through the Center for the Science and Engineering of Materials at the California Institute of Technology. P.R. acknowledges the hospitality of the Graduate Aerospace Laboratories at the California Institute of Technology (GALCIT).
\bibliography{WeakComprRefs_jkn}

\begin{thebibliography}{10}

\bibitem{2discher}
Dennis~E Discher, Paul Janmey, and Yu-li Wang.
\newblock Tissue cells feel and respond to the stiffness of their substrate.
\newblock {\em Science}, 310(5751):1139--1143, 2005.

\bibitem{5lo}
Chun-Min Lo, Hong-Bei Wang, Micah Dembo, and Yu-li Wang.
\newblock Cell movement is guided by the rigidity of the substrate.
\newblock {\em Biophysical Journal}, 79(1):144--152, 2000.

\bibitem{6reinhart}
Cynthia~A Reinhart-King, Micah Dembo, and Daniel~A Hammer.
\newblock Cell-cell mechanical communication through compliant substrates.
\newblock {\em Biophysical Journal}, 95(12):6044--6051, 2008.

\bibitem{9winer}
Jessamine~P Winer, Shaina Oake, and Paul~A Janmey.
\newblock Non-linear elasticity of extracellular matrices enables contractile
  cells to communicate local position and orientation.
\newblock {\em PLoS One}, 4(7):e6382, 2009.

\bibitem{shi2014}
Quanming Shi, Rajarshi~P Ghosh, Hanna Engelke, Chris~H Rycroft, Luke Cassereau,
  James~A Sethian, Valerie~M Weaver, and Jan~T Liphardt.
\newblock Rapid disorganization of mechanically interacting systems of mammary
  acini.
\newblock {\em Proceedings of the National Academy of Sciences},
  111(2):658--663, 2014.

\bibitem{he}
Shijie He, Yewang Su, Baohua Ji, and Huajian Gao.
\newblock Some basic questions on mechanosensing in cell--substrate
  interaction.
\newblock {\em Journal of the Mechanics and Physics of Solids}, 70:116--135,
  2014.

\bibitem{1vogel}
Viola Vogel and Michael Sheetz.
\newblock Local force and geometry sensing regulate cell functions.
\newblock {\em Nature Reviews Molecular Cell Biology}, 7(4):265--275, 2006.

\bibitem{notbth}
Jacob Notbohm.
\newblock {\em Dynamics of Cell-Matrix Mechanical Interactions in Three
  Dimensions}.
\newblock PhD thesis, California Institute of Technology, 2013.

\bibitem{notb}
Jacob Notbohm, Ayelet Lesman, Phoebus Rosakis, David~A Tirrell, and Guruswami
  Ravichandran.
\newblock Microbuckling of fibrin provides a mechanism for cell mechanosensing.
\newblock {\em Journal of The Royal Society Interface}, 12(108):20150320, 2015.

\bibitem{26franck}
C~Franck, S~Hong, SA~Maskarinec, DA~Tirrell, and G~Ravichandran.
\newblock Three-dimensional full-field measurements of large deformations in
  soft materials using confocal microscopy and digital volume correlation.
\newblock {\em Experimental Mechanics}, 47(3):427--438, 2007.

\bibitem{25rudnicki}
Mathilda~S Rudnicki, Heather~A Cirka, Maziar Aghvami, Edward~A Sander, Qi~Wen,
  and Kristen~L Billiar.
\newblock Nonlinear strain stiffening is not sufficient to explain how far
  cells can feel on fibrous protein gels.
\newblock {\em Biophysical Journal}, 105(1):11--20, 2013.

\bibitem{Lakes}
R~Lakes, P~Rosakis, and A~Ruina.
\newblock Microbuckling instability in elastomeric cellular solids.
\newblock {\em Journal of Materials Science}, 28(17):4667--4672, 1993.

\bibitem{Kim}
Oleg~V. Kim, Rustem~I. Litvinov, John~W. Weisel, and Mark~S. Alber.
\newblock Structural basis for the nonlinear mechanics of fibrin networks under
  compression.
\newblock {\em Biomaterials}, 35(25):6739 -- 6749, 2014.

\bibitem{kim2}
Oleg~V Kim, Xiaojun Liang, Rustem~I Litvinov, John~W Weisel, Mark~S Alber, and
  Prashant~K Purohit.
\newblock Foam-like compression behavior of fibrin networks.
\newblock {\em Biomechanics and modeling in mechanobiology}, pages 1--16, 2015.

\bibitem{piech}
Izabela~K Piechocka, Rommel~G Bacabac, Max Potters, Fred~C MacKintosh, and
  Gijsje~H Koenderink.
\newblock Structural hierarchy governs fibrin gel mechanics.
\newblock {\em Biophysical journal}, 98(10):2281--2289, 2010.

\bibitem{11janmey}
Paul~A Janmey, Margaret~E McCormick, Sebastian Rammensee, Jennifer~L Leight,
  Penelope~C Georges, and Fred~C MacKintosh.
\newblock Negative normal stress in semiflexible biopolymer gels.
\newblock {\em Nature Materials}, 6(1):48--51, 2007.

\bibitem{conti}
Enrico Conti and Fred~C MacKintosh.
\newblock Cross-linked networks of stiff filaments exhibit negative normal
  stress.
\newblock {\em Physical review letters}, 102(8):088102, 2009.

\bibitem{poynting}
JH~Poynting.
\newblock On pressure perpendicular to the shear planes in finite pure shears,
  and on the lengthening of loaded wires when twisted.
\newblock {\em Proceedings of the Royal Society of London. Series A, Containing
  Papers of a Mathematical and Physical Character}, 82(557):546--559, 1909.

\bibitem{mihai}
L~Angela Mihai and Alain Goriely.
\newblock Positive or negative poynting effect? the role of adscititious
  inequalities in hyperelastic materials.
\newblock {\em Proceedings of the Royal Society A: Mathematical, Physical and
  Engineering Science}, 467(2136):3633--3646, 2011.

\bibitem{10storm}
Cornelis Storm, Jennifer~J Pastore, Fred~C MacKintosh, Tom~C Lubensky, and
  Paul~A Janmey.
\newblock Nonlinear elasticity in biological gels.
\newblock {\em Nature}, 435(7039):191--194, 2005.

\bibitem{roeder}
Blayne~A Roeder, Klod Kokini, Jennifer~E Sturgis, J~Paul Robinson, and Sherry~L
  Voytik-Harbin.
\newblock Tensile mechanical properties of three-dimensional type i collagen
  extracellular matrices with varied microstructure.
\newblock {\em Journal of biomechanical engineering}, 124(2):214--222, 2002.

\bibitem{12brown}
A.E.X. Brown, R.I. Litvinov, D.E. Discher, P.K. Purohit, and J.W. Weisel.
\newblock Multiscale mechanics of fibrin polymer: Gel stretching with protein
  unfolding and loss of water.
\newblock {\em Science}, 325(5941):741--744, 2009.

\bibitem{jawerth}
Albert~James Licup, Stefan M{\"u}nster, Abhinav Sharma, Michael Sheinman,
  Louise~M Jawerth, Ben Fabry, David~A Weitz, and Fred~C MacKintosh.
\newblock Stress controls the mechanics of collagen networks.
\newblock {\em arXiv preprint arXiv:1503.00924}, 2015.

\bibitem{16shokef}
Yair Shokef and Samuel~A Safran.
\newblock Scaling laws for the response of nonlinear elastic media with
  implications for cell mechanics.
\newblock {\em Physical Review Letters}, 108(17):178103, 2012.

\bibitem{shenoy}
Hailong Wang, AS~Abhilash, Christopher~S Chen, Rebecca~G Wells, and Vivek~B
  Shenoy.
\newblock Long-range force transmission in fibrous matrices enabled by
  tension-driven alignment of fibers.
\newblock {\em Biophysical journal}, 107(11):2592--2603, 2014.

\bibitem{notbles}
Jacob Notbohm, Ayelet Lesman, David~A Tirrell, and Guruswami Ravichandran.
\newblock Quantifying cell-induced matrix deformation in three dimensions based
  on imaging matrix fibers.
\newblock {\em Integrative Biology}, 2015.

\bibitem{wolf}
Katarina Wolf and Peter Friedl.
\newblock Extracellular matrix determinants of proteolytic and non-proteolytic
  cell migration.
\newblock {\em Trends in cell biology}, 21(12):736--744, 2011.

\end{thebibliography}

\end{document}